# On the observability of Majoron emitting double beta decays

M. Hirsch, H.V. Klapdor–Kleingrothaus, S.G. Kovalenko, [1]
H. Päs

*Max-Planck-Institut für Kernphysik, P.O.Box 10 39 80, D-69029 Heidelberg, Germany*

**Abstract**

Because of the fine–tuning problem in classical Majoron models in recent years several new models were invented. It is pointed out that double beta decays with Majoron emission depend on new matrix elements, which have not been considered in the literature. A calculation of these matrix elements and phase space integrals is presented. We find that for new Majoron models extremly small decay rates are expected. *PACS* 13.15;23.40;21.60E;14.80

In many theories of physics beyond the standard model neutrinoless double beta decays can occur with the emission of new bosons, so–called Majorons [1–4]:

$$2n \rightarrow 2p + 2e^{-} + \phi \tag{1}$$

$$2n \rightarrow 2p + 2e^{-} + 2\phi. \tag{2}$$

Since classical Majoron models [1,5] require severe fine–tuning in order to preserve existing bounds on neutrino masses and at the same time get an observable rate for Majoron emitting double beta decays in recent years several new Majoron models have been constructed [6–8], where the terminus Majoron means in a more general sense light or massless bosons with couplings to neutrinos. The main novel features of the "New Majorons" are that they can carry units of leptonic charge, that there can be Majorons which are no Goldstone bosons [6] and that decays with the emission of two Majorons [4,7] can occur. The latter can be scalar–mediated or fermion–mediated. In vector Majoron models the Majoron becomes the longitudinal component of

[1] on leave from *Joint Institute for Nuclear Research, Dubna, Russia*

| case | modus | Goldstone boson | L | n | Matrix element |
|---|---|---|---|---|---|
| IB | $\beta\beta\phi$ | no | 0 | 1 | $M_F - M_{GT}$ |
| IC | $\beta\beta\phi$ | yes | 0 | 1 | $M_F - M_{GT}$ |
| ID | $\beta\beta\phi\phi$ | no | 0 | 3 | $M_{F\omega^2} - M_{GT\omega^2}$ |
| IE | $\beta\beta\phi\phi$ | yes | 0 | 3 | $M_{F\omega^2} - M_{GT\omega^2}$ |
| IIB | $\beta\beta\phi$ | no | -2 | 1 | $M_F - M_{GT}$ |
| IIC | $\beta\beta\phi$ | yes | -2 | 3 | $M_{CR}$ |
| IID | $\beta\beta\phi\phi$ | no | -1 | 3 | $M_{F\omega^2} - M_{GT\omega^2}$ |
| IIE | $\beta\beta\phi\phi$ | yes | -1 | 7 | $M_{F\omega^2} - M_{GT\omega^2}$ |
| IIF | $\beta\beta\phi$ | Gauge boson | -2 | 3 | $M_{CR}$ |

Table 1
Different Majoron models according to Bamert/Burgess/Mohapatra[9]. The case IIF corresponds to the model of Carone[10].

a massive gauge boson [8] emitted in double beta processes. For simplicity we will call it Majoron, too.
In tab. 1 the nine Majoron models we considered are summarized. [7,8] It is divided in the sections I for lepton number breaking and II for lepton number conserving models. The table shows also whether the corresponding double beta decay is accompanied by the emission of one or two Majorons.
The next three entries list the main features of the models: The third column lists whether the Majoron is a Goldstone boson or not (or a gauge boson in case of vector Majorons IIF). In column four the leptonic charge L is given. In column five the "spectral index" $n$ of the sum energy of the emitted electrons is listed, which is defined from the phase space of the emitted particles, $G \sim (Q_{\beta\beta} - T)^n$, where $Q_{\beta\beta}$ is the energy release of the decay and $T$ the sum energy of the two electrons. The different shapes can be used to discriminate the different decay modes from each other and the double beta decay with emission of two neutrinos. In the last column we listed the nuclear matrix elements which will be defined in more detail later. Nuclear matrix elements are necessary to convert half lives (or limits thereof) into values for the effective Majoron–neutrino coupling constant, using the approximate (see below) relations: [4,9]

$$[T_{1/2}]^{-1} = |<g_\alpha>|^m \cdot |M_\alpha|^2 \cdot G_{BB_\alpha} \quad (3)$$

with $m = 2$ for $\beta\beta\phi$-decays or $m = 4$ for $\beta\beta\phi\phi$–decays. The index $\alpha$ in eq. (3) indicates that effective coupling constants $g_\alpha$, nuclear matrix elements $M_\alpha$ and phase spaces $G_{BB_\alpha}$ differ for different models.
As shown in tab. 1, several Majoron models with different theoretical motivation can lead to signals in double beta decays which are experimentally indistinguishable. The interpretation of experimental half life limits in terms of the "effective Majoron–neutrino coupling constant" is therefore model dependent. Subsequently we give a brief summary of the theoretical background on which our conclusions on the different Majoron models are based.

Single Majoron emitting double beta decays ($0\nu\beta\beta\phi$) can be roughly divided into two classes, $n = 1$ (case IB, IC and IIB) and $n = 3$ (IIC and IIF) decays. As has been noted in [7] as long as $0\nu\beta\beta$ decay has not been observed, the three $n = 1$ decays are indistinguishable from each other. We will call these Majorons "ordinary", since they contain the subgroup IC, which leads to the classical Majoron models. [1,2,10] For all ordinary Majorons the effective Majoron–neutrino interaction Lagrangian, leading to $0\nu\beta\beta\phi$ decay is [2,6]

$$L^{O.M.}_{\phi\nu\nu} = -\frac{1}{2}\overline{\nu_i}(a_{ij}P_L + b_{ij}P_R)\nu_j\phi^* + h.c. \qquad (4)$$

Here, $P_{R/L} = 1/2(1 \pm \gamma_5)$. Using eq. (4) the amplitude corresponding to the Feynman graph is, in the notation of [6]

$$\mathcal{A}^{O.M.}(0\nu\beta\beta\phi) = 4\sqrt{2}\sum_{i,j} V_{ei}V_{ej}\int \frac{d^4q}{(2\pi)^4}\frac{m_i m_j a_{ij} + q^2 b_{ij}}{(q^2 - m_i^2 + i\epsilon)(q^2 - m_j^2 + i\epsilon)}(w_F - w_{GT}) \qquad (5)$$

$V_{ei}$, $V_{ej}$ are elements of the neutrino mixing matrix, $m_i$ and $m_j$ denote neutrino mass eigenvalues and $w_{F/GT}$ are nuclear matrix elements containing double Fermi and Gamow–Teller operators. To arrive at the factorized decay rate eq. (3), the usual assumption $m_{i,j} \ll q \approx p_F \approx \mathcal{O}(100MeV)$, where $p_F$ is the typical Fermi momentum of nucleons, is made. By this assumption the term proportional to $a_{ij}$ can be dropped and the effective coupling constant is defined as:

$$\langle g\rangle^{O.M.} = \sum_{i,j} V_{ei}V_{ej}b_{ij}. \qquad (6)$$

In this approximation matrix elements for ordinary Majoron decays coincide with the leading terms $M_{GT}$ and $M_F$ of the well–known mass mechanism of $0\nu\beta\beta$ decay.

Burgess and Cline advocated the so–called charged Majoron model IIC. [6] In this model the effective interaction Lagrangian is

$$L^{C.M.}_{\phi\nu\nu} = -\frac{i}{2f}\overline{\nu}\gamma^\mu(A_LP_L + A_RP_R)\nu\partial_\mu\phi + h.c. \qquad (7)$$

Note, that in the charged Majoron model the two additional powers of $n$ in the phase space integrals originate from the derivative coupling of the Majoron in $L^{C.M.}_{\phi\nu\nu}$. As shown in [6], for charged Majorons the contribution from the leading order matrix elements to the decay rate vanishes identically, so that one has to go to the next higher order in the non–relativistic impulse approximation of hadronic currents. The amplitude for $0\nu\beta\beta\phi$ decay is then given by

$$\mathcal{A}^{C.M.}(0\nu\beta\beta\phi) = 8\sqrt{2}\sum_{i,j} V_{ei}V_{ej} \int \frac{d^4q}{(2\pi)^4} \frac{q^2 b_{ij}}{(q^2 - m_i^2 + i\epsilon)(q^2 - m_j^2 + i\epsilon)}(w_5 + w_6) \quad (8)$$

which leads to an effective coupling constant $\langle g\rangle^{C.M.}$ as in the ordinary Majoron case, but with $b_{ij}$ given by $b = \frac{i}{f}(A_L m^* + m^* A_R)$, with the neutrino mass matrix $m$, generator matrices $A_{L/R}$ and the decay constant $f$. The hadronic term $w_6$ is similar but not identical to the recoil matrix element of $0\nu\beta\beta$ decay induced by right–handed currents. This difference has turned out to be important. In the notation of [6]

$$w_5 = \frac{i\mu\vec{q}}{|q|^2(q_0^2 - \mu^2 + i\epsilon)}\langle F|e^{-i\vec{q}\vec{r}}[g_A^2(C_n\vec{\sigma_m} - C_m\vec{\sigma_n}) + g_V^2(\mathbf{D}_m - \mathbf{D}_n)]|I\rangle, \quad (9)$$

$$w_6 = \frac{\mu g_A g_V \vec{q}}{|q|^2(q_0^2 - \mu^2 + i\epsilon)}\langle F|e^{-i\vec{q}\vec{r}}[\mathbf{D}_n \times \vec{\sigma_m} + \mathbf{D}_m \times \vec{\sigma_n}]|I\rangle, \quad (10)$$

in which the summation over $\sum_{nm} \tau_n^+\tau_m^+$ is suppressed. Here, $C_n$ and $\mathbf{D}_n$ are nuclear recoil terms [9]

$$C_n = (\mathbf{P}_n + \mathbf{P}'_n)\cdot\vec{\sigma_n}/(2M_n) - (E_n - E'_n)(\mathbf{P}_n - \mathbf{P}'_n)\cdot\vec{\sigma_n}/(2m_\pi^2) \quad (11)$$

$$\mathbf{D}_n = [(\mathbf{P}_n + \mathbf{P}'_n) + i\mu_\beta(\mathbf{P}_n - \mathbf{P}'_n)\times\vec{\sigma_n}]/(2M_n) \quad (12)$$

$\mathbf{P}_n$ ($E_n$) and $\mathbf{P}'_n$ ($E'_n$) are momenta (energies) of initial and final state nucleons, $m_\pi$ is the pion and $M_n$ the nucleon mass and $\mu_\beta$ originates from the weak magnetism.

The terms of $w_5$ are neglected compared to $w_6$ due to the estimation $(\mathbf{P}_n + \mathbf{P}'_n) \leq (\mathbf{P}_n - \mathbf{P}'_n), (E_n - E'_n) \leq \mathcal{O}(Q_{\beta\beta})$. [9] Following [11] we will also keep only the central part of the recoil term $\mathbf{D}$. Although both are approximations, which needs to be checked numerically, we do not expect it to affect any of our conclusions.

Finally for vector Majoron models (case IIF) [8]

$$L_{\phi\nu\nu}^{V.M.} = -\frac{i}{2f}\overline{\nu}\gamma^\mu(c_{ij}P_L + d_{ij}P_R)\nu X^\mu + h.c. \quad (13)$$

where $X^\mu$ is the emitted massive gauge boson. The effective coupling constant can be defined as in the ordinary Majoron model, with the replacement $b_{ij} = \frac{1}{2M}(c_{ij}m_j - m_i d_{ij})$, where $M$ is the gauge boson mass. As discussed in [8], the vector Majoron amplitude approaches the charged Majoron one in the limit of vanishing gauge boson masses, which we assume in the phase space integration. They depend on the same nuclear matrix elements than the charged Majoron discussed above. We will therefore not repeat the definitions here.

Double Majoron emitting decays ($0\nu\beta\beta\phi\phi$), mediated by fermions, can have either spectral index $n = 7$ or $n = 3$, depending on whether the Majoron couples derivatively suppressed or not. [7]

In addition, in principle $0\nu\beta\beta\phi\phi$ decays could also be mediated by exotic scalars. The amplitude of scalar–mediated decays, however, is expected to be very much suppressed, since the scalars must have masses larger than about 50 GeV due to the LEP–measurements. [7] We will therefore concentrate on the fermion–mediated decays.
The Yukawa coupling of the Majoron to the neutrinos for the $n = 3$ decays (cases ID, IE, IID) is given as:

$$L^{yuk}_{\phi\nu\nu} = -\overline{\nu_i}(A_{ia}P_L + B_{ia}P_R)N_a\phi + h.c. \tag{14}$$

where $A_{ia}$ and $B_{ia}$ represent arbitrary Yukawa–coupling matrices and $N_a$ are sterile neutrinos. The corresponding amplitude for $0\nu\beta\beta\phi\phi$ decay is

$$\mathcal{A}^{D.M.}(0\nu\beta\beta\phi\phi) = \sqrt{\frac{2}{3\pi^2}}\sum_{i,j,a} V_{ei}V_{ej}\int\frac{d^4q}{(2\pi)^4} \times \frac{\mathcal{N}_{ija}}{(q^2-m_i^2+i\epsilon)(q^2-m_j^2+i\epsilon)(q^2-m_a^2+i\epsilon)}(w_F - w_{GT}). \tag{15}$$

Although for $\mathcal{A}^{D.M.}(0\nu\beta\beta\phi\phi)$ the same combination of nuclear operators appears $(w_F - w_{GT})$, note the additional $(q^2-m^2)^{-1}$ compared to $\mathcal{A}^{O.M.}(0\nu\beta\beta\phi)$. $\mathcal{N}_{ija}$ in (15) is given by

$$\mathcal{N}_{ija} = -q^2(A_{ia}B_{ja}m_{\nu_i} + A_{ja}B_{ia}m_{\nu_j} + B_{ia}B_{ja}m_{N_a}) + A_{ia}A_{ja}m_{\nu_i}m_{\nu_j}m_{N_a} \tag{16}$$

In order to separate the particle physics parameters from the nuclear structure calculation, it is most convenient to neglect the last term in eq. (16). This can be justified by considering that the mass eigenvalues $m_{\nu_{i,j}} \ll p_F$ so that the last term in eq. (16) for not too large $m_{N_a}$ is suppressed compared to the first three by at least $m_{\nu_{i,j}}/p_F \simeq \mathcal{O}(10^{-5-6})$. Then, the $q^2$ is absorbed into the neutrino potentials and we redefine $\mathcal{N}_{ija}$ to obtain the effective coupling constant as

$$\langle g\rangle = (\frac{1}{m_e}\sum_{ij} V_{ei}V_{ej}[A_{ia}B_{ja}m_{\nu_i} + A_{ja}B_{ia}m_{\nu_j} + B_{ia}B_{ja}m_{N_a}])^{\frac{1}{2}} \tag{17}$$

Note, that we have arbitrarily absorbed a factor of $m_e^{-1}$ into the definition of $\langle g\rangle$ here to get for the effective coupling a dimensionless quantity.
For the $n = 7$ $0\nu\beta\beta\phi\phi$ decays, the effective Lagrangian is (IIE/fermion mediated):

$$L_{\phi\nu\nu} = -i\overline{\nu_i}\gamma^\mu(X_{ia}P_L + Y_{ia}P_R)N_a\partial_\mu\phi + h.c. \tag{18}$$

Again, $N_a$ denotes a sterile neutrino and the derivative coupling of $\phi$ accounts for the additional powers of $n$ in the phase space integrals. The amplitude for $n = 7$ decays is the same as for the $n = 3$ case, discussed above, with the replacement: $\mathcal{N}_{ija} = X_{ia}Y_{ia}m_{N_a}$.

Note, that $X_{ia}$ and $Y_{ia}$ have the dimension of an inverse mass. Therefore, also $\langle g \rangle$ has a dimension of an inverse mass. To define a dimensionless coupling constant in this case one would have to specify the symmetry breaking scale, which is however undetermined by the model.
For the Majoron models considered in this work there are five nuclear matrix elements to be calculated. Within the closure approximation they are defined as:

$$M_F = (\frac{g_V^2}{g_A^2}) < N_f \| h_{mass}(\mu, \vec{r}) \tau_n^+ \tau_m^+ \| N_i > \tag{19}$$

$$M_{GT} =< N_f \| h_{mass}(\mu, \vec{r}) \tau_n^+ \tau_m^+ \vec{\sigma}_n \vec{\sigma}_m \| N_i > \tag{20}$$

$$M_{CR} = (\frac{g_V}{g_A}) \frac{f_W}{3} < N_f \| h_R(\mu, \vec{r}) \tau_n^+ \tau_m^+ \vec{\sigma}_n \vec{\sigma}_m \| N_i > \tag{21}$$

$$M_{F\omega^2} = (\frac{g_V^2}{g_A^2}) < N_f \| h_{\omega^2}(\mu, \vec{r}) \tau_n^+ \tau_m^+ \| N_i > \tag{22}$$

$$M_{GT\omega^2} =< N_f \| h_{\omega^2}(\mu, \vec{r}) \tau_n^+ \tau_m^+ \vec{\sigma}_n \vec{\sigma}_m \| N_i > \tag{23}$$

where $h_\alpha$ denote the neutrino potentials

$$h_{mass}(\mu, \vec{r}) = \frac{R}{2\pi^2} \int \frac{d^3\vec{q}}{\omega} \frac{e^{i\vec{q}\vec{r}}}{\omega + \mu} \tag{24}$$

$$h_R(\mu, \vec{r}) = \frac{1}{4\pi^2} (\frac{1}{M}) \int \frac{d^3\vec{q}}{\omega} e^{i\vec{q}\vec{r}} \frac{\mu + 2\omega}{(\mu + \omega)^2} \tag{25}$$

$$h_{\omega^2}(\mu, \vec{r}) = \frac{m_e^2 R}{16\pi^2} \int d^3\vec{q} q^2 e^{i\vec{q}\vec{r}} \frac{3\mu^2 + 9\mu\omega + 8\omega^2}{\omega^5 (\mu + \omega)^3} \tag{26}$$

Here $\mu = \langle E_N - E_I \rangle$ denotes the average excitation energy of the intermediate nuclear states. $\omega = \sqrt{q^2 + m^2}$ is the energy of the neutrino and since we assume all neutrinos to be light, the indices on neutrino masses have been dropped. Note, that in order to define matrix elements dimensionless we follow the convention of [9]. That is $h_{mass}(\vec{r})$ and $h_{\omega^2}(\vec{r})$ are arbitrarily multiplied by the nuclear radius $R = r_0 A^{\frac{1}{3}}$ with $r_0 = 1.2$ fm, while $h_R(\vec{r})$ includes the nucleon mass. Compensating factors appear in the prefactors of the phase space integrals.
We have carried out a numerical calculation of these matrix elements within the pn–QRPA model of [12,13]. To estimate the uncertainties of the nuclear structure matrix elements the parameter dependence of the numerical results has been investigated. Since the matrix elements $M_{GT}$ and $M_F$ have been studied before, [12] we will concentrate on $M_{CR}$, $M_{GT\omega^2}$ and $M_{F\omega^2}$. $M_{GT}$ and $M_F$ can be calculated with an accuracy of about a factor of 2[12].

The matrix element $M_{CR}$ shows a very similar behaviour as $M_{GT}$. This is in agreement with the expectation, since only the central part of the recoil terms is taken into account, so that apart from the different neutrino potential $M_{CR}$ has the same structure as $M_{GT}$. Neither variations of the strength of the particle–particle force $g_{pp}$ nor a change in the intermediate state energies significantly affects the numerical value of $M_{CR}$. We therefore conclude that $M_{CR}$ should be accurate up to a factor of 2, as is expected for $M_{GT}$.
Unfortunately, in the case of the matrix elements $M_{GT\omega^2}$ and $M_{F\omega^2}$ the situation is very different. Both, variations of $g_{pp}$ or $\mu$, can change the numerical results drastically (fig. 1). In fact, it is found that $M_{GT\omega^2}$ displays a very similar dependence on $g_{pp}$ as has been reported in pn–QRPA studies of $2\nu\beta\beta$ decay matrix elements. [12] Especially important is that in the region of the most probable value of $g_{pp}$ $M_{GT\omega^2}$ crosses zero.
Also for variations of the assumed average intermediate state energy a rather strong dependence of the results on the adopted value of $\mu$ has been found. As a consequence of this unpleasant strong dependence, for an accurate prediction of $M_{GT\omega^2}$ and $M_{F\omega^2}$ it seems necessary to go beyond the closure approximation.
The basic reason for the unusual sensitivity of $M_{GT\omega^2}$ and $M_{F\omega^2}$ on $\mu$ can be traced back to a certain difference in the neutrino potential of these matrix elements compared to $M_{GT/F}$, $h_{mass}(\mu,\vec{r}) \sim \omega^{-2}$ while $h_{\omega^2}(\mu,\vec{r}) \sim \omega^{-4}$. Contributions from very low momenta are therefore much preferred in $h_{\omega^2}(\mu,\vec{r})$ compared to $h_{mass}(\mu,\vec{r})$. (Note, that this leads also to a much smaller value for a typical $\omega$ than the naive expectation of $\omega \sim p_F \sim \mathcal{O}(50-100)$ MeV!). With typical $\omega$ of only $\mathcal{O}$(few) MeV the strong dependence of $h_{\omega^2}(\mu,\vec{r})$ on $\mu$ becomes obvious.
Results of the calculation for various experimentally interesting isotopes are summarized in table 2. Note that the matrix elements are valid for the limit of small intermediate particle masses, up to the order of 10 MeV. If any of the virtual particles in the Feynman graphs can have masses larger than 10 MeV, the matrix elements are no longer constant and the values in table 2 should only be taken as upper limits for the analysis of data.
In comparison to the nuclear matrix elements phase space integrals can be calculated very accurately, so uncertainties of this calculation will not be discussed. We define the phase space integral as

$$G_{BB_\alpha} = a_\alpha \cdot \int (Q_{\beta\beta} - \epsilon_1 - \epsilon_2)^n \prod_k p_k \epsilon_k f(\epsilon_k) d\epsilon_k \qquad (27)$$

where the prefactor $a_\alpha$ depends on the Majoron mode under consideration. A summary of the definitions is given in table 3. $Q_{\beta\beta}$ is the maximum decay energy, $\epsilon_k$ and $p_k$ are the energies and momenta of the outgoing electrons and $f(\epsilon_k)$ is the Fermi function calculated according to the description of. [9] Note the large difference in the phase space values of the old ($n = 1$) and new Majoron models.

| nucleus | $M_F - M_{GT}$ | $M_{CR}$ | $M_{F\omega^2} - M_{GT\omega^2}$ |
|---|---|---|---|
| Ge 76 | 4.33 | 0.16 | $\sim 10^{-3\pm1}$ |
| Se 82 | 4.03 | 0.14 | $\sim 10^{-3\pm1}$ |
| Mo 100 | 4.86 | 0.16 | $\sim 10^{-3\pm1}$ |
| Cd 116 | 3.29 | 0.10 | $\sim 10^{-3\pm1}$ |
| Te 128 | 4.49 | 0.14 | $\sim 10^{-3\pm1}$ |
| Te 130 | 3.90 | 0.12 | $\sim 10^{-3\pm1}$ |
| Xe 136 | 1.82 | 0.05 | $\sim 10^{-3\pm1}$ |
| Nd 150 | 5.29 | 0.15 | $\sim 10^{-3\pm1}$ |

Table 2
Dimensionless nuclear matrix elements of Majoron emitting modes calculated in this work

| nucleus | $\beta\beta\phi$ n=1 | $\beta\beta\phi$ n=3 | $\beta\beta\phi\phi$ n=3 | $\beta\beta\phi\phi$ n=7 |
|---|---|---|---|---|
| $a_\alpha$ | $\frac{(G_F g_A)^4 \cdot 2 \cdot m_e^2}{256\pi^7 ln(2)\hbar(m_e R)^2}$ | $\frac{(G_F g_A)^4 \cdot 2}{64\pi^7 ln(2)\hbar}$ | $\frac{(G_F g_A)^4 \cdot 2}{12288\pi^9 ln(2)\hbar(m_e R)^2}$ | $\frac{(G_F g_A)^4 \cdot 2}{215040\pi^9 m_e^4 ln(2)\hbar(m_e R)^2}$ |
| Ge 76 | $1.25 \cdot 10^{-16}$ | $2.07 \cdot 10^{-19}$ | $6.32 \cdot 10^{-19}$ | $1.21 \cdot 10^{-18}$ |
| Se 82 | $1.03 \cdot 10^{-15}$ | $3.49 \cdot 10^{-18}$ | $1.01 \cdot 10^{-17}$ | $7.73 \cdot 10^{-17}$ |
| Mo 100 | $1.80 \cdot 10^{-15}$ | $7.28 \cdot 10^{-18}$ | $1.85 \cdot 10^{-17}$ | $1.54 \cdot 10^{-16}$ |
| Cd 116 | $1.75 \cdot 10^{-15}$ | $6.95 \cdot 10^{-18}$ | $1.60 \cdot 10^{-17}$ | $1.03 \cdot 10^{-16}$ |
| Te 128 | $1.02 \cdot 10^{-17}$ | $5.96 \cdot 10^{-21}$ | $1.28 \cdot 10^{-20}$ | $1.20 \cdot 10^{-21}$ |
| Te 130 | $1.35 \cdot 10^{-15}$ | $4.97 \cdot 10^{-18}$ | $1.06 \cdot 10^{-17}$ | $4.83 \cdot 10^{-17}$ |
| Xe 136 | $1.40 \cdot 10^{-15}$ | $5.15 \cdot 10^{-18}$ | $1.06 \cdot 10^{-17}$ | $4.54 \cdot 10^{-17}$ |
| Nd 150 | $1.07 \cdot 10^{-14}$ | $7.27 \cdot 10^{-17}$ | $1.41 \cdot 10^{-16}$ | $1.85 \cdot 10^{-15}$ |

Table 3
Values of phase space integrals calculated in this work

Having calculated nuclear matrix elements and phase space integrals, it is straightforward to derive limits on the effective Majoron–neutrino coupling constants for the various Majoron models from experiment.
Although experimental half life limits are comparable for all decay modes, as observed recently for $^{76}$Ge decay [14,15], restrictive limits on the coupling constants of ordinary Majoron models contrast with limits on any of the new Majoron models, which will be weaker by (3–4) orders of magnitude.
The surprisingly weak limits which one obtains for the neutrino–Majoron coupling constant due to small matrix elements and phase spaces for all of the new Majoron models, require further explanation. (Note that the follow-ing discussion is independent of the isotope under consideration.) Consider, for example, ordinary and charged Majoron $0\nu\beta\beta\phi$ decays. Limits on the effective coupling constant for single Majoron emitting decays will scale as $\langle g\rangle \sim \mathcal{A}^{-1}(T_{1/2}\cdot G_{BB})^{-\frac{1}{2}}$. Thus, the relative sensitivity of a double beta decay experiment on ordinary and charged Majoron decays can be expressed as $\frac{\langle g\rangle^{O.M.}}{\langle g\rangle^{C.M.}} \sim \frac{\mathcal{A}^{C.M.}}{\mathcal{A}^{O.M.}}\left(\frac{T_{1/2}^{C.M.}}{T_{1/2}^{O.M.}}\right)^{\frac{1}{2}}\cdot(Q_{\beta\beta}-T)$.
Inserting the definitions of the corresponding amplitudes, it is clear that even if the half life limit derived for the charged Majoron decay equals that of the ordinary Majoron mode, limits on the charged Majoron–neutrino coupling constant will be weaker by $M_n/(Q_{\beta\beta}-T)\simeq 1000$ ! (Note, that this crude estimation is to first approximation independent of nuclear structure properties.)
A similar analysis can be easily done for double Majoron emitting decays. Again, very crudely, a reduced sensitivity of $(48\pi^2)\cdot p_F/(Q_{\beta\beta}-T)\simeq$ (few) $\times 10^4$ for $n=3$ double Majoron decay, compared to ordinary Majoron decays, is expected. Here, the factor $(48\pi^2)$ is due to the phase space integration over the additional emitted particle, while the latter factor comes from the additional propagator.
One might think that since our definition of the effective coupling constant for the $n=3$ $0\nu\beta\beta\phi\phi$ decays includes a factor $m_{N_a}/m_e$, where $m_{N_a}$ is the sterile neutrino mass, one could get $\langle g\rangle$ easily as large as wanted, since the mass of the sterile neutrino is not bounded experimentally. However, matrix elements will fall off $M \sim m_{N_a}^{-2}$ as soon as $m_{N_a}$ is larger than the typical momenta. While for the matrix elements $M_{GT/F}$ for ordinary Majoron decays such a reduction occurs starting from masses of exchanged virtual particles in the region of $100-1000$ MeV, for $M_{GT\omega^2/F\omega^2}$ the suppression will be important already for much smaller masses (see the $E_n$–dependance fig. 1).
Since the sensitivity of double beta decay experiments to the new Majoron models is so weak, it might be interesting to compare expected half lives for the different models for different $\langle g\rangle$, $\langle g\rangle \approx 10^{-4}$ as a typical sensitivity in coupling constant for ordinary Majoron models and $\langle g\rangle = 1$ as an upper possible limit allowed by perturbation theory, with current experimental limits of $\mathcal{O}(10^{22})$ years (see tab. 4). From this consideration it is very unlikely that any of the new Majoron models can produce an observable rate in planned or ongoing double beta decay experiments. Only the charged and the vector Majoron model [6,8] could produce an observable effect if $\sum_{ij} V_{ei}V_{ej}$ is not smaller than 0.1 and the real coupling constant of order $\mathcal{O}(1)$.

## Acknowledgements

The authors would like to thank C.P. Burgess and E. Takasugi for several discussions on the theoretical aspects of Majoron models. The research described in this publication was made possible in part (M.H.) by the Deutsche

| model | $T_{1/2}(<g>=10^{-4})$ | $T_{1/2}(<g>=1)$ | $T_{1/2exp}$ |
|---|---|---|---|
| IB,IC,IIB | $4 \cdot 10^{22}$ | $4 \cdot 10^{14}$ | $1.67 \cdot 10^{22}$ |
| ID,IE,IID | $10^{38-42}$ | $10^{22-26}$ | $1.67 \cdot 10^{22}$ |
| IIC,IIF | $2 \cdot 10^{28}$ | $2 \cdot 10^{20}$ | $1.67 \cdot 10^{22}$ |
| IIE | $10^{38-42}$ | $10^{22-26}$ | $3.37 \cdot 10^{22}$ |

Table 4
Comparison of half lives calculated for different $<g>$–values for the new Majoron models with experimental best fit values[16,18]

Forschungsgemeinschaft (446 JAP–113/101/0 and Kl 253/8–1) and (S.G.K.) by Grant No. RFM300 from the International Science Foundation.

$M_{GT\omega^2} - M_{F\omega^2}$

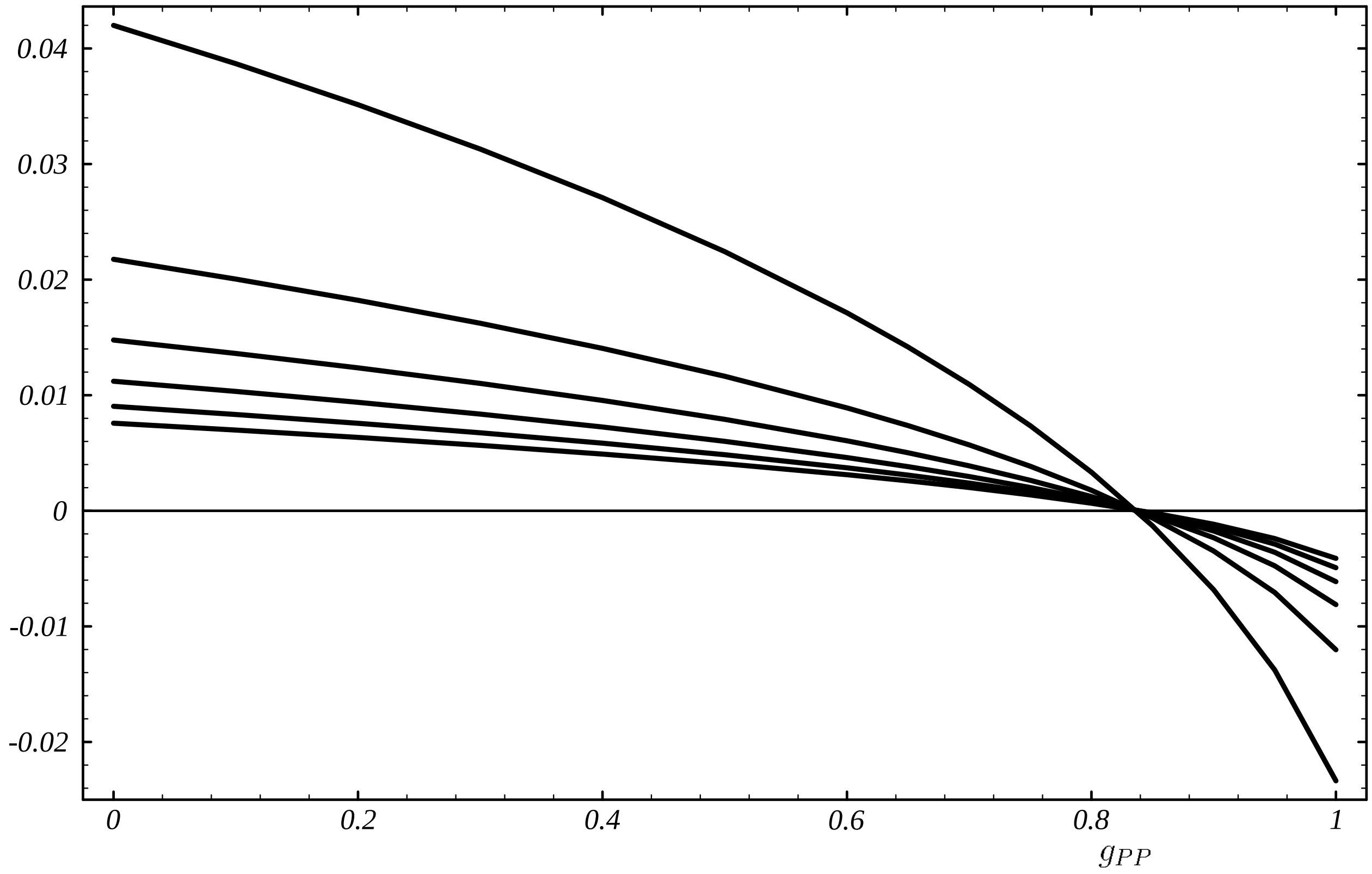

Fig. 1. $M_{GT\omega^2} - M_{F\omega^2}$ dependence of $g_{PP}$ for different intermediate state energies $E_n$ =4 (top on the left),8,12,16,20,24 (bottom on the left) MeV for $^{76}Ge$